\documentclass[aps,prd,twocolumn,groupedaddress,showpacs,superscriptaddress,floatfix,nofootinbib,10pt]{revtex4-1}

\usepackage{amsmath}\allowdisplaybreaks[1]
\usepackage{graphicx}
\usepackage{amssymb}
\usepackage{bm}
\usepackage{array}
\usepackage[colorlinks,linkcolor=blue,urlcolor=blue,anchorcolor=blue,citecolor=blue]{hyperref}
\usepackage{slashed}
\usepackage{braket}
\usepackage{xcolor}
\usepackage{tabularx}
\usepackage{comment}

\begin{document}

\title{Experimental constraint on quark electric dipole moments}

\newcommand*{\DUKE}{Department of Physics, Duke University and Triangle Universities Nuclear Laboratory, Durham, North Carolina 27708, USA}\affiliation{\DUKE}
\newcommand*{\DKU}{Duke Kunshan University, Kunshan, Jiangsu 215316, China}\affiliation{\DKU}

\author{Tianbo Liu}
\email{liutb@jlab.org}
\affiliation{\DUKE}
\affiliation{\DKU}

\author{Zhiwen Zhao}
\affiliation{\DUKE}

\author{Haiyan Gao}
\affiliation{\DUKE}
\affiliation{\DKU}


\begin{abstract}

 The electric dipole moments (EDMs) of nucleons are sensitive probes of additional $\cal CP$ violation sources beyond the standard model to account for the baryon number asymmetry of the universe. As a fundamental quantity of the nucleon structure, tensor charge is also a bridge that relates nucleon EDMs to quark EDMs. With a combination of nucleon EDM measurements and tensor charge extractions, we investigate the experimental constraint on quark EDMs, and its sensitivity to $\cal CP$ violation sources from new physics beyond the electroweak scale. We obtain the current limits on quark EDMs as $1.27\times10^{-24}\,e\cdot{\rm cm}$ for the up quark and $1.17\times10^{-24}\,e\cdot{\rm cm}$ for the down quark at the scale of $4\,\rm GeV^2$. We also study the impact of future nucleon EDM and tensor charge measurements, and show that upcoming new experiments will improve the constraint on quark EDMs by about three orders of magnitude leading to a much more sensitive probe of new physics models.
\end{abstract}


\maketitle

\section{Introduction}

Symmetries play a central role in physics. Discrete symmetries, charge conjugate ($\cal C$), parity ($\cal P$), and time-reversal ($\cal T$), were believed to be conserved until the discovery of parity violation in weak interactions suggested by Lee and Yang~\cite{Lee:1956qn} and then confirmed in nuclear beta decays~\cite{Wu:1957my} and successive meson decays~\cite{Garwin:1957hc}. It is a cornerstone of the standard model (SM) of particle physics. Subsequently, the violation of the combination of charge conjugate and parity, $\cal CP$, was discovered in neutral kaon decays~\cite{Christenson:1964fg} and also observed in $B$-meson and strange $B$-meson decays in recent years~\cite{Aubert:2001nu,Abe:2001xe,Aaij:2013iua}. The $\cal CP$ violation is one of Sakharov conditions for generating a baryon number asymmetry from a symmetric initial state~\cite{Sakharov:1967rr}. Although the Kobayash-Maskawa (KM) mechanism~\cite{Kobayashi:1973fv} provides a consistent and economical SM description of all observed $\cal CP$ violating phenomena in collider physics, the $\cal CP$ violating phase in the Cabibbo-Kobayashi-Maskawa (CKM) matrix is not enough to account for the observed asymmetry of matter and antimatter in the visible universe. Therefore, additional $\cal CP$ violation sources are required, unless one accepts the fine tuning solution of the initial condition prior to the Big Bang.

A permanent electric dipole moment (EDM) of any particle with a non-degenerate ground state violates both parity and time-reversal symmetries. Assuming $\cal CPT$ invariance, a consequence of local quantum field theories with Lorentz invariance~\cite{Luders:1954zz,Lueders:1992dq,Bell:1996nh,Pauli:1955}, it is a signal of $\cal CP$ violation.
As the CKM complex phase requires the participation of three fermion generations, the EDM of light quarks is highly suppressed by the flavor changing interactions at the three-loop level, and hence the KM mechanism only results in an extremely small EDM~\cite{Czarnecki:1997bu,Gavela:1981sk,Khriplovich:1981ca}. Therefore, the quark EDM is one of the most sensitive probes to new physics beyond the SM. 

Due to the confinement property of strong interaction, the quark EDM is not directly measurable, but instead one can derive it from nucleon EDM measurements. A bridge that relates the quark EDM and the nucleon EDM is the tensor charge, which is a fundamental QCD quantity defined by the matrix element of the tensor current. It also represents the transverse spin carried by the quarks in a transversely polarized nucleon in the parton model. The double role of the tensor charge in the understanding of strong interaction and the search for new physics received great interest in recent years~\cite{Courtoy:2015haa}. Nowadays, the progress on lattice QCD gives a better control on the uncertainty of the tensor charge calculation~\cite{Bhattacharya:2015esa}. On the experimental side, the proposed experiments at the 12\,GeV upgraded Jefferson Lab will have an unprecedented precision in measurements of the tensor charge~\cite{Gao:2010av}.

The history of nucleon EDM experiments can be traced back to the 1950s, and the first experiment was proposed by Purcell and Ramsey  using the neutron-beam magnetic resonance method~\cite{Purcell:1950zz,Smith:1957ht}. The current upper limit on the neutron EDM from direct measurements is $3.0\times10^{-26}\,e\cdot{\rm cm}$ (90\% C.L.)~\cite{Afach:2015sja}, which was obtained with ultra-cold neutrons permeated in uniform electric and magnetic fields by measuring the difference of Larmor precession frequencies when flipping the electric field,
\begin{equation}
h\nu=|2\mu_nB \pm 2d_nE|.
\end{equation}
The current upper limit on the proton EDM is derived from the mercury atomic EDM limit with the Schiff moment method~\cite{Dmitriev:2003sc}, because the existence of an electric monopole usually overwhelms the dipole signal. The most recent measurement of the EDM of $^{199}$Hg atoms sets the upper limit on the mercury atomic EDM to $7.4\times10^{-30}\,e\cdot{\rm cm}$ (95\% C.L.)~\cite{Graner:2016ses}. Following the Schiff moment method~\cite{Dmitriev:2003sc} and the theoretical calculations~\cite{Ban:2010ea}, we obtain the upper limit on the proton EDM as $2.6\times10^{-25}\,e\cdot{\rm cm}$~\footnote{The derived upper limit on the proton EDM in Ref.~\cite{Graner:2016ses} is $2.0\times10^{-25}\,e\cdot{\rm cm}$, which was obtained with the relation from the random-phase approximation with core polarization~\cite{Dmitriev:2003sc}. Here we use the fully self-consistent calculations in Ref.~~\cite{Ban:2010ea} and include a theoretical uncertainty to account for the difference among interaction models.}. As mentioned in Ref.~\cite{Dmitriev:2003sc}, the uncertainty in this derivation is not statistical and we cannot give the probability distribution. As a conservative estimation, we expect the derived value of the proton EDM limit is no worse than a 90\% confidence level.

The effective Lagrangian that contributes to the nucleon EDM can be expressed up to dimension-six as~\cite{Grzadkowski:2010es}
\begin{widetext}
\begin{equation}
\begin{split}
\mathcal{L}_{\rm eff}&=
-\bar{\theta}\frac{g_s^2}{64\pi^2}\epsilon^{\mu\nu\rho\sigma}G_{\mu\nu}^a G_{\rho\sigma}^a
-\frac{1}{2}\sum_q d_q\bar{\psi}_q i\sigma^{\mu\nu}\gamma_5\psi_q F_{\mu\nu}
-\frac{1}{2}\sum_q \tilde{d}_q\bar{\psi}_q i\sigma^{\mu\nu}t^a\psi_q G_{\mu\nu}^a\\
&\quad +\frac{1}{6}d_W f^{abc}\epsilon^{\mu\nu\rho\sigma}G_{\mu\nu}^a G_{\rho\lambda}^b G_{\sigma}^{c\,\lambda}
+\sum_{i,j,k,l}C_{ijkl}\bar{\psi}_i\Gamma\psi_j \bar{\psi}_k\Gamma'\psi_l,
\end{split}\label{Leff}
\end{equation}
\end{widetext}
where $g_s$ is the strong coupling constant, $G_{\mu\nu}^a$ is the gluon field, $F_{\mu\nu}$ is the electromagnetic field, and $\psi_q$ is the quark field. The first term, a dimension-four operator, is allowed in the standard model as the QCD $\theta$-term, where the overall phase of the quark mass matrix is absorbed into $\bar{\theta}$. It could in principle generate large hadronic EDMs, but the upper limit on the neutron EDM constrains the coefficient to $|\bar{\theta}|\leq 10^{-10}$. The two dimension-five terms are respectively the quark EDM $d_q$ and the quark chromoelectric dipole moment $\tilde{d}_q$. In order to restore the $SU(2)\times U(1)$ symmetry above the electroweak scale, a Higgs field insertion should be included in these two terms~\cite{DeRujula:1990db}. Therefore, they are essentially dimension-six operators, and are often in practice supplied by an insertion of the quark mass as $m_q/\Lambda^2$, where $\Lambda$ represents a large mass scale. For consistency, other dimension-six operators, the three-gluon Weinberg operator and the four-fermion interactions, should also be introduced.

In this paper, focusing on the quark EDM term, we present the experimental limit on quark EDMs with the combination of nucleon EDM measurements and tensor charge extractions, and the impact of the next generation EDM experiments and the planned precision measurements of the tensor charge. The constraint on new physics is also discussed.

\section{Tensor charge and quark EDM}

\begin{figure*}[ht]
\centering
\includegraphics[width=0.52\textwidth]{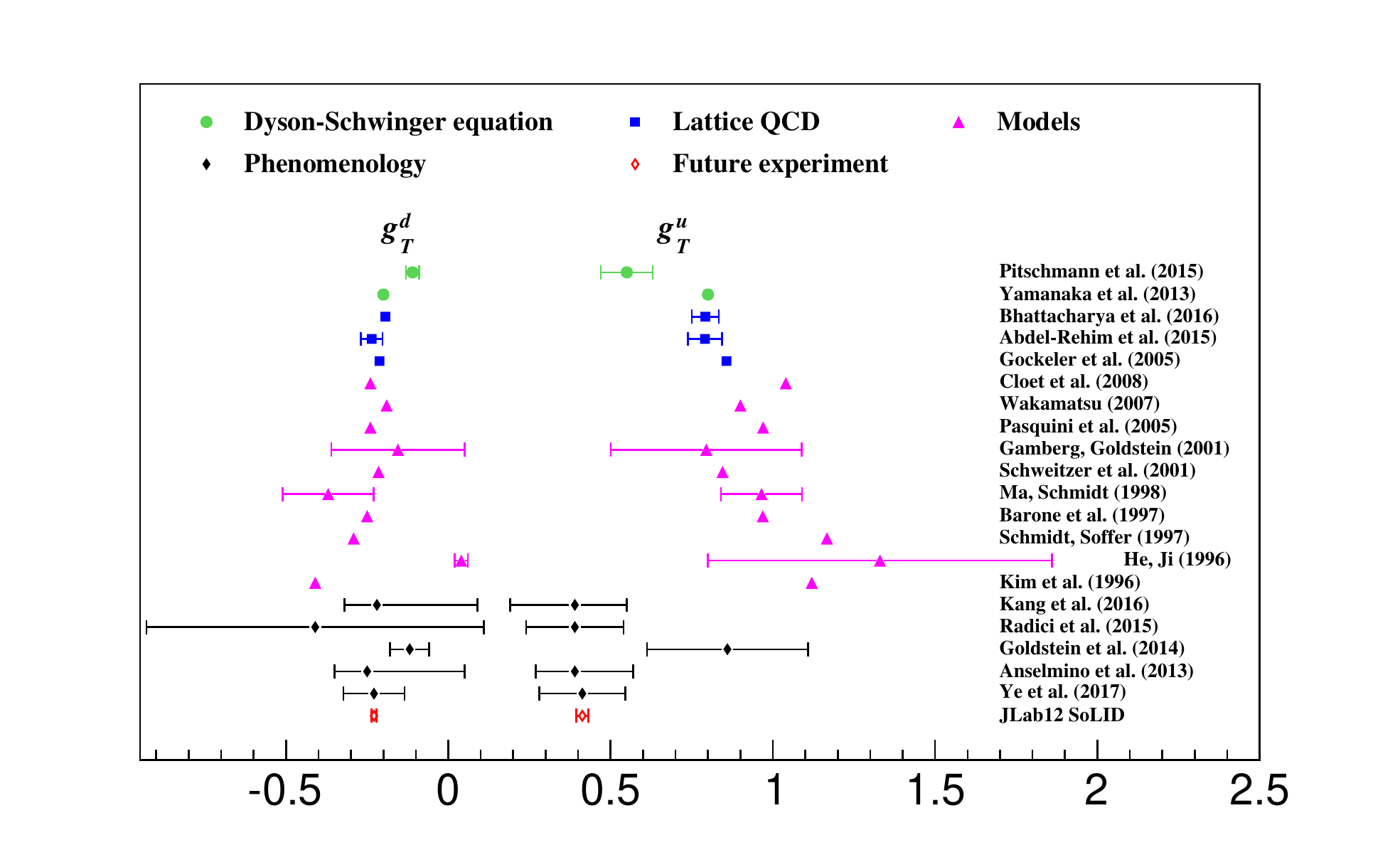}
\includegraphics[width=0.473\textwidth]{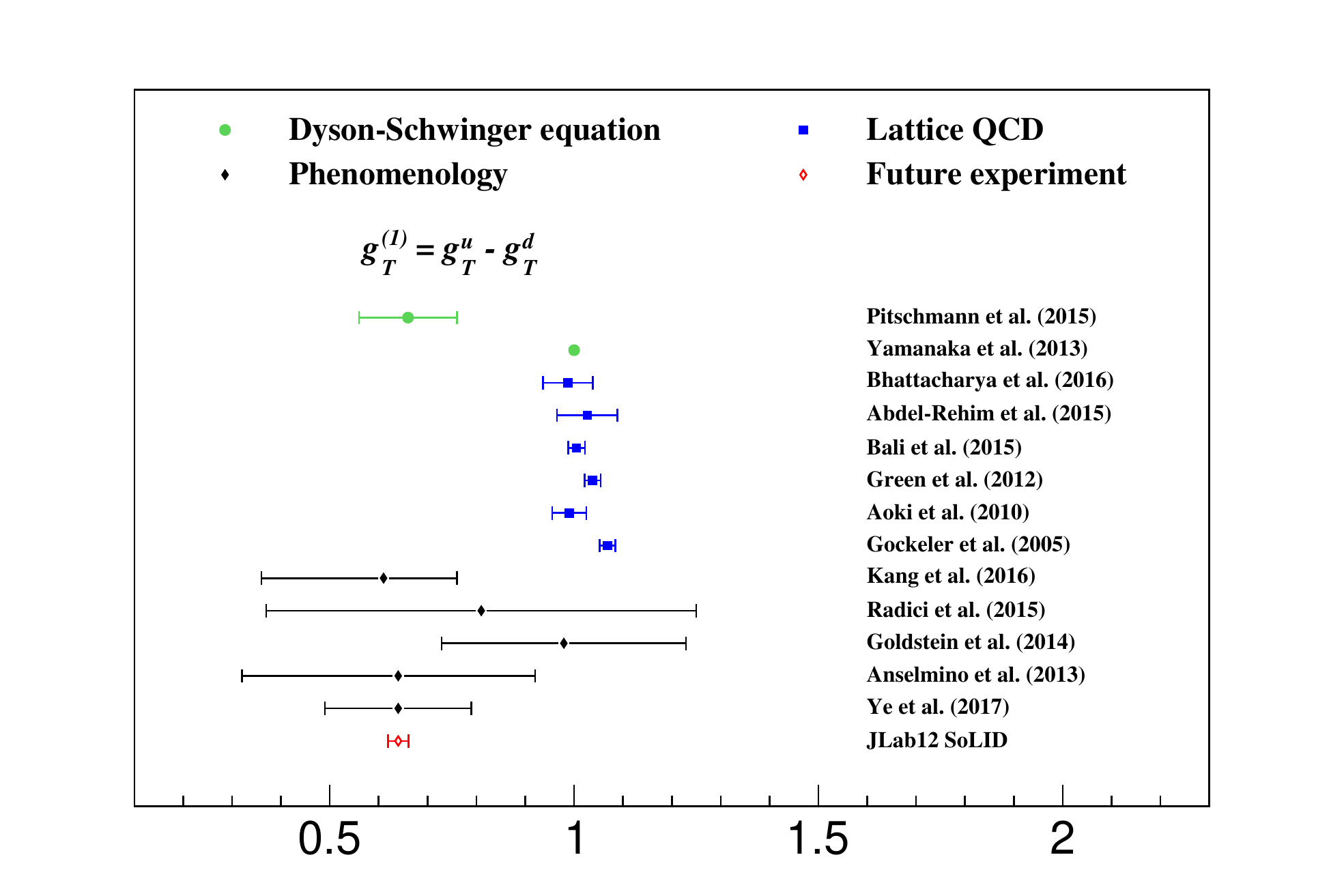}
\caption{Tensor charge results. The left panel shows the values of $u$ and $d$ quark tensor charges, and the right panel shows the values of the isovector tensor charge. The round (green) points are from Dyson-Schwinger equation calculations~\cite{Pitschmann:2014jxa,Yamanaka:2013zoa}, the square (blue) points are from lattice QCD calculations~\cite{Bhattacharya:2016zcn,Abdel-Rehim:2015owa,Gockeler:2005cj,Bali:2014nma,Green:2012ej,Aoki:2010xg}, the triangle (magenta) points are from model calculations~\cite{Cloet:2007em,Wakamatsu:2007nc,Pasquini:2005dk,Gamberg:2001qc,Schweitzer:2001sr,Ma:1997pm,Barone:1996un,Schmidt:1997vm,He:1996wy,Kim:1996vk}, the filled diamond (black) points are phenomenological extractions from data~\cite{Kang:2015msa,Radici:2015mwa,Goldstein:2014aja,Anselmino:2013vqa,Ye:2016prn}, and the hollow diamond (red) points are the projection of JLab-12GeV SoLID experiments based on the most recent global analysis~\cite{Ye:2016prn}. The results quoted from the references are evaluated at different scales as explained in the text.}\label{tensor_charge}
\end{figure*}

The nucleon EDM is related to the quark EDM as~\cite{Ellis:1996dg,Bhattacharya:2012bf,Pitschmann:2014jxa}
\begin{align}
d_p&=g_T^u\,d_u+g_T^d\,d_d+g_T^s\,d_s,\label{pedmtc}\\
d_n&=g_T^d\,d_u+g_T^u\,d_d+g_T^s\,d_s,\label{nedmtc}
\end{align}
where the isospin symmetry is applied in Eq.~\eqref{nedmtc}. In this study we neglect heavy flavor contributions. The coefficient $g_T^{u,d,s}$ is the tensor charge, which is defined by the matrix element of a local operator as
\begin{equation}
\langle p,\sigma|\bar{\psi}_q i\sigma^{\mu\nu}\psi_q|p,\sigma\rangle=g_T^{q}\,\bar{u}(p,\sigma)i\sigma^{\mu\nu}u(p,\sigma).
\end{equation}
In the naive nonrelativistic quark model, it can be obtained from the $SU(6)$ spin-flavor wave function~\cite{Gursey:1992dc}
\begin{equation}
\Psi_p=\frac{1}{\sqrt{18}}(2u_\uparrow u_\uparrow d_\downarrow - u_\uparrow u_\downarrow d_\uparrow - u_\downarrow u_\uparrow d_\uparrow + {\rm permutations})
\end{equation}
as
\begin{equation}
g_T^{u}=\frac{4}{3},\quad g_T^{d}=-\frac{1}{3}.
\end{equation}
Due to relativistic effects, the tensor charge values reduce from the prediction of the naive quark model~\cite{Ma:1997gy}, and differ from the axial-vector charge, which is defined by the matrix element of the axial-vector current. As shown in Figure~\ref{tensor_charge}, the tensor charge has been calculated in many phenomenological models~\cite{Cloet:2007em,Wakamatsu:2007nc,Pasquini:2005dk,Gamberg:2001qc,Schweitzer:2001sr,Ma:1997pm,Barone:1996un,Schmidt:1997vm,He:1996wy,Kim:1996vk}, and with some nonperturbative methods, such as Dyson-Schwinger equation calculations~\cite{Pitschmann:2014jxa,Yamanaka:2013zoa} and lattice QCD simulations~\cite{Bhattacharya:2016zcn,Abdel-Rehim:2015owa,Gockeler:2005cj,Bali:2014nma,Green:2012ej,Aoki:2010xg}.

In the quark-parton model, the tensor charge is equal to the first moment of quark transversity distribution,
\begin{equation}
g_T^{q}=\int_0^1 dx[h_1^q(x)-h_1^{\bar{q}}(x)],
\end{equation}
where $x$ represents the longitudinal momentum fraction carried by the quark. The transversity distribution $h_1(x)$ as a leading-twist parton distribution function is interpreted as the net density of transversely polarized quarks in a transversely polarized proton. Unlike its longitudinal counter part, the helicity distribution, which measures the density of longitudinally polarized quarks in a longitudinally polarized proton, the transversity distribution is a chiral-odd quantity, which results in a simpler QCD evolution effect without mixing with gluons and as such is dominated by valence quarks~\cite{Barone:2001sp}. However, the chiral-odd property makes it decouple at the leading-twist from the inclusive deep-inelastic scattering (DIS) process, which is usually the most efficient approach to measure parton distributions, and hence it should be measured by coupling to another chiral-odd quantity. The semi-inclusive DIS (SIDIS) is one of the processes that can be used to measure transversity distributions. At the leading twist, the transversity distribution can be extracted from a transverse target single spin asymmetry, the Collins asymmetry~\cite{Collins:1992kk}, which arises from the convolution of the transversity distribution and the Collins fragmentation function within the framework of the transverse momentum dependent (TMD) factorization. It can also be measured in the collinear factorization through the dihadron process~\cite{Bacchetta:2011ip} by coupling to the dihadron fragmentation function. Besides, the tensor charge can be estimated from generalized parton distribution (GPD) extractions, because the transversity distribution is the forward limit of a chiral-odd GPD. The tensor charge values from some recent global analyses are shown in Figure~\ref{tensor_charge}. The values of Anselmino {\it et al.}~\cite{Anselmino:2013vqa}, Kang {\it et al.}~\cite{Kang:2015msa}, and Ye {\it et al.}~\cite{Ye:2016prn} are from the global fit of transversity TMDs. The TMD evolution effect is not taken into account in Anselmino {\it et al.}~\cite{Anselmino:2013vqa}, while it is included in the other two. The values of Radici {\it et al.}~\cite{Radici:2015mwa} are from the analysis of the dihadron process within the collinear factorization. The values of Goldstein {\it et al.}~\cite{Goldstein:2014aja} are from the GPD extractions by analyzing mesons ($\pi^0$ and $\eta$) exclusive electroproduction data. Within the large uncertainties, the results from different groups are consistent with each other.

We should note that the tensor charge is scale dependent, and it follows the evolution equation at the leading order as~\cite{Artru:1989zv}
\begin{equation}
g_T^q(Q^2)=g_T^q(Q_0^2)\left[\frac{\alpha_s(Q^2)}{\alpha_s(Q_0^2)}\right]^{\frac{4}{33-2n_f}},
\end{equation}
where $n_f$ is the number of flavors. The results in Figure~\ref{tensor_charge} are at different scales. The Dyson-Schwinger equation results of Pitschmann {\it et al.}~\cite{Pitschmann:2014jxa} and Yamanaka {\it et al.}~\cite{Yamanaka:2013zoa} and the lattice simulation results of Bhattacharya {\it et al.}~\cite{Bhattacharya:2016zcn}, Abdel-Rehim {\it et al.}~\cite{Abdel-Rehim:2015owa}, Gockeler {\it et al.}~\cite{Gockeler:2005cj}, Bali {\it et al.}~\cite{Bali:2014nma}, Green {\it et al.}~\cite{Green:2012ej}, and Aoki {\it et al.}~\cite{Aoki:2010xg} are at $4\,\rm GeV^2$. The model calculation results of Cloet {\it et al.}~\cite{Cloet:2007em} are at $0.16\,\rm GeV^2$, the results of Wakamatsu~\cite{Wakamatsu:2007nc} are at $0.36\,\rm GeV^2$, the results of Pasquini {\it et al.}~\cite{Pasquini:2005dk} are at $0.079\,\rm GeV^2$, the results of Gamberg and Goldstein~\cite{Gamberg:2001qc} are at $1\,\rm GeV^2$, the results of Schweitzer {\it et al.}~\cite{Schweitzer:2001sr} are at $0.36\,\rm GeV^2$, the results of Ma and Schmidt~\cite{Ma:1997pm} are at $3\sim10\,\rm GeV^2$, the result of Barone {\it et al.}~\cite{Barone:1996un} are at $25\,\rm GeV^2$, the results of Schmidt and Soffer~\cite{Schmidt:1997vm} are at $0.09\,\rm GeV^2$, the results of He and Ji~\cite{He:1996wy} are at $1\,\rm GeV^2$, and the results of Kim {\it et al.}~\cite{Kim:1996vk} are at $0.36\,\rm GeV^2$. The phenomenological extraction results of Kang {\it et al.}~\cite{Kang:2015msa} are at $10\,\rm GeV^2$, the results of Radici {\it et al.}~\cite{Radici:2015mwa} are at $1\,\rm GeV^2$, the results of Goldstein {\it et al.}~\cite{Goldstein:2014aja} are at $4\,\rm GeV^2$, the results of Anselmino {\it et al.}~\cite{Anselmino:2013vqa} are at $0.8\,\rm GeV^2$, and the results of Ye {\it et al.} and the SoLID projections~\cite{Ye:2016prn} are at $2.4\,\rm GeV^2$. To see the size of the evolution effect, we quote here the results at two different scales in Ref.~\cite{Ye:2016prn} as 
\begin{align}
g_T^u=0.413\pm0.133,\quad g_T^d=-0.229\pm0.094,
\end{align}
at $2.4\,\rm GeV^2$, and
\begin{align}
g_T^u=0.395\pm0.128,\quad g_T^d=-0.219\pm0.090,
\end{align}
at $10\,\rm GeV^2$.

As seen from Eqs.~\eqref{pedmtc} and~\eqref{nedmtc}, the constraint on quark EDMs can be obtained from the knowledge of tensor charges and the nucleon EDM measurements. To have quantitative estimations, we take the upper limit on the proton EDM derived from the most recent measurement of $^{199}$Hg EDM~\cite{Graner:2016ses} with the Schiff moment method
\begin{equation}
|d_p|\leq 2.6\times10^{-25}\,e\cdot{\rm cm},\label{current_dp}
\end{equation}
and the upper limit on the neutron EDM from the direct measurement with ultra-cold neutrons~\cite{Afach:2015sja}
\begin{equation}
|d_n|\leq 3.0\times10^{-26}\,e\cdot{\rm cm}.\label{current_dn}
\end{equation}
For the tensor charge, we use the results from the global analysis in Ref.~\cite{Ye:2016prn}, which includes the TMD evolution effect. At the scale of $4\,\rm GeV^2$, the extracted tensor charges for up and down quarks are 
\begin{equation}\label{current_gt}
g_T^{u}=0.405\pm0.130,\quad
g_T^{d}=-0.225\pm0.092,
\end{equation}
with the uncertainties given at 90\% confidence level. In this analysis the strange quark transversity and thus its tensor charge are set to zero. Then the strange quark contribution to nucleon EDMs vanishes. However, since the quark EDM is expected to be proportional to the quark mass for a large class of models in which the Lagrangian shares the same form for each fermion family member, the contribution from the strange quark term could be large even with a small tensor charge value~\cite{Bhattacharya:2015esa}. To account for the uncertainty from the strange quark contribution, we take the strange quark tensor charge value from the recent lattice simulation~\cite{Bhattacharya:2015esa},
\begin{equation}
g_T^s=0.008\pm0.009,
\end{equation}
at the scale of $4\,\rm GeV^2$. The uncertainty is understood as one standard deviation ($1\sigma$). For consistency, we multiplies it by a factor of $1.65$, which corresponds to a 90\% C.L. based on the normal distribution assumption. Following the method in~\cite{Bhattacharya:2015esa}, we relate $d_s$ and $d_d$ with the quark mass ratio $m_s/m_d$. In Ref.~\cite{Bhattacharya:2015esa} the ratio is chosen as 20, and in Ref.~\cite{Olive:2016xmw} the ratio is evaluated as $17\sim22$. Since we are estimating the upper limit on quark EDMs, we use the value $m_s/m_d=22$ in our analysis to maximize the uncertainty from the strange quark term.

\begin{figure*}[ht]
\centering
\includegraphics[width=0.49\textwidth]{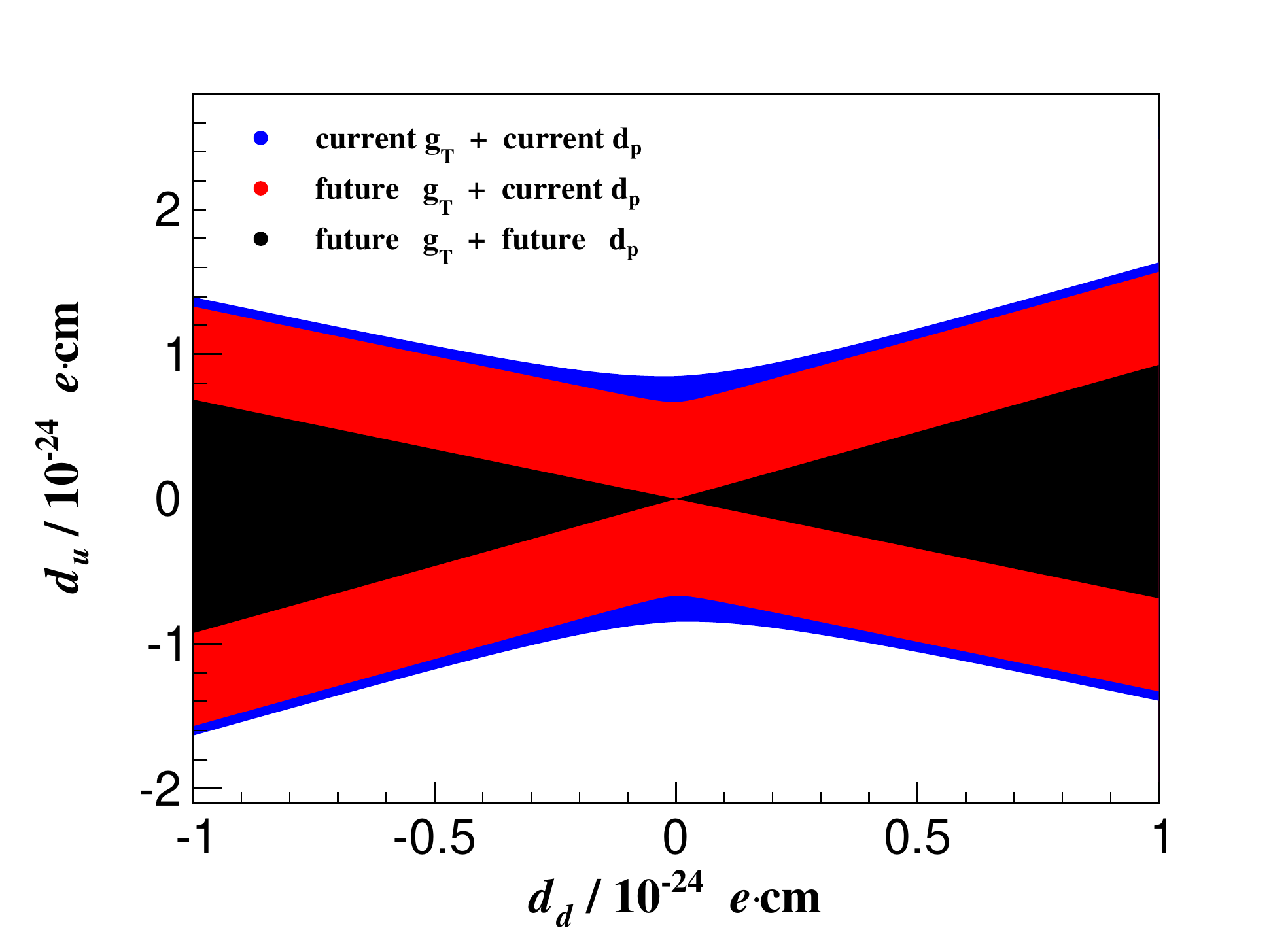}
\includegraphics[width=0.49\textwidth]{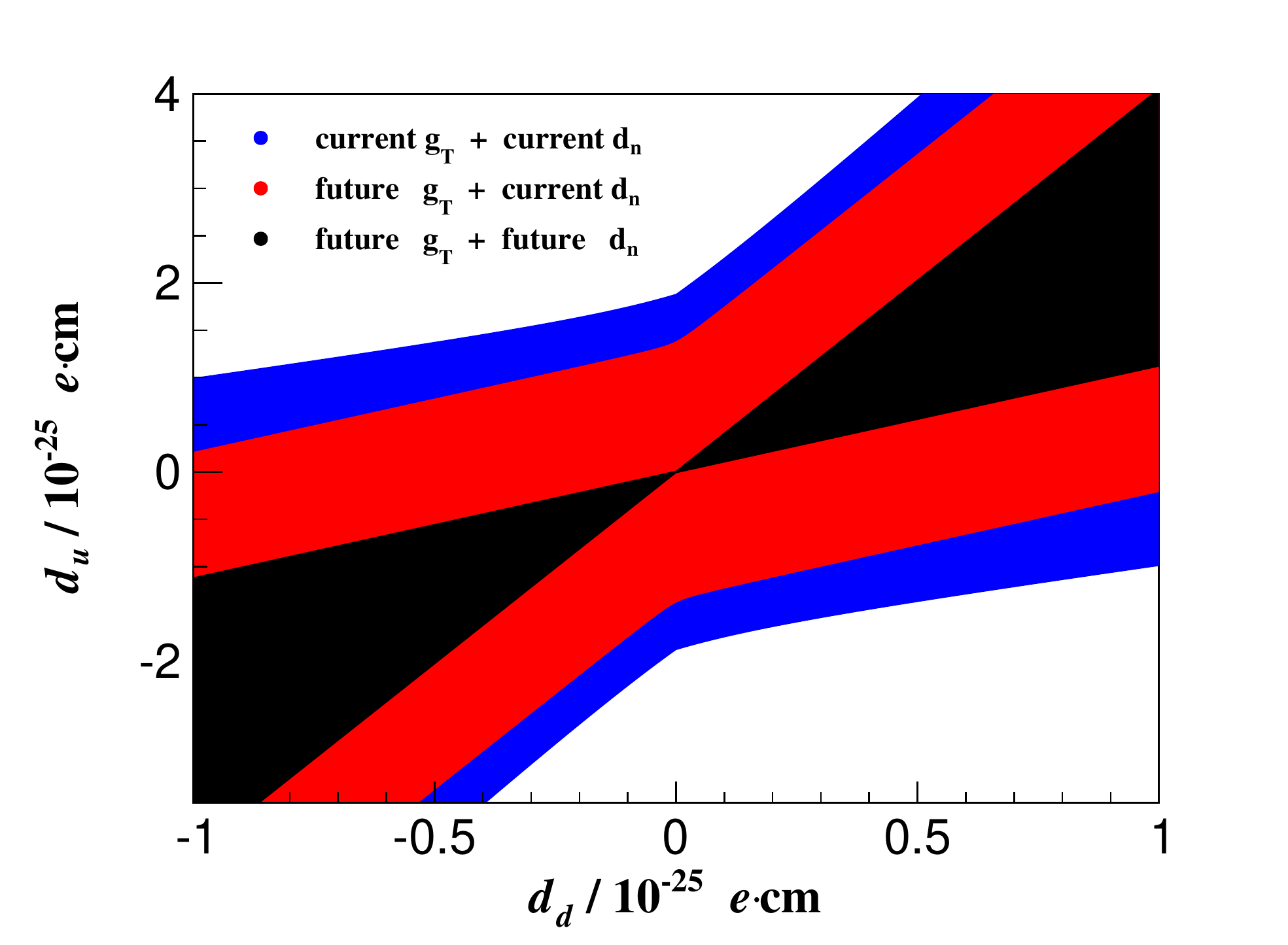}
\caption{Constraints on quark EDMs with the upper limits on nucleon EDMs and the tensor charge extractions. The left panel shows the constraints by the upper limit on the proton EDM, and the right panel shows the constraints by the upper limit on the neutron EDM. The current tensor charge precision refers to $g_T^{u}=0.405\pm0.130$ and $g_T^{d}=-0.225\pm0.092$, and the future tensor charge precision refers to $g_T^u=0.405\pm0.018$ and $g_T^d=-0.225\pm0.008$~\cite{Ye:2016prn}. The strange quark tensor charge $g_T^{s}=0.008\pm0.015$~\cite{Bhattacharya:2015esa} is used for both current and future tensor charge precisions. The current nucleon EDM limit refers to $|d_p|\leq2.6\times10^{-25}\,e\cdot{\rm cm}$ derived from mercury atomic EDM measurement~\cite{Graner:2016ses} and $|d_n|\leq3.0\times10^{-26}\,e\cdot{\rm cm}$ from neutron EDM measurement~\cite{Afach:2015sja}, and the future nucleon EDM limit means $|d_p|\leq2.6\times10^{-29}\,e\cdot{\rm cm}$ and $|d_n|\leq3.0\times10^{-28}\,e\cdot{\rm cm}$. The constraints are understood at the scale of $4\,\rm GeV^2$.}\label{nucleon_edm}
\end{figure*}

With the proton and neutron EDM limits and the tensor charge values above, we obtain the constraint on quark EDMs in Figure~\ref{nucleon_edm}. For flavor separation, we combine the results from proton and neutron EDM limits and obtain the constraint on up and down quark EDMs as
\begin{align}
|d_u|&\leq1.15\times10^{-24}\,e\cdot{\rm cm},\\
|d_d|&\leq1.06\times10^{-24}\,e\cdot{\rm cm}.
\end{align}
Since all scale dependent quantities in the analysis are evaluated at $4\,\rm GeV^2$, these constraints should be understood at the same scale. The confidence level of the constraints is 90\%, because the nucleon EDM limits and the tensor charge uncertainties applied in the evaluation are at 90\% confidence level. We should note that the isospin symmetry, which will bring in additional uncertainties, is applied in our analysis to perform flavor separation. With our current knowledge, it is hard to quantify this uncertainty. Hence we simply add a 10\% uncertainty to account for the isospin symmetry breaking effect, and the final results of up and down quark EDM limits are listed in Table~\ref{quark_edm}. Our current constraint on light quark EDMs is at $10^{-24}\,e\cdot{\rm cm}$ level.

\section{The impact of future experiments}

The constraint on quark EDMs is affected by both the sensitivity of nucleon EDM measurements and the precision of tensor charge extractions. To improve the quark EDM limit and thus its sensitivity to new physics, we need efforts from both sides. In this section, we estimate the impact of the planned experiments in the next ten years.

One of the main goals of the 12 GeV upgraded CEBAF Jefferson Lab is to understand the partonic structure of the nucleon. The SIDIS experiments at Jefferson Lab, particularly those with SoLID~\cite{Chen:2014psa} which will combine large acceptance and high luminosities, are aiming to have an unprecedented precision in measurements of quark three dimensional distributions in the momentum space. The Collins asymmetry, a transverse target single spin asymmetry, is one of the highlighted measurements aiming to extract quark transversity distributions and thus the tensor charge. A quantitative study~\cite{Ye:2016prn} shows that SoLID SIDIS experiments will improve the precision of tensor charge extractions by one order of magnitude. The projection based on the global analysis~\cite{Ye:2016prn} gives
\begin{align}
g_T^u=0.405\pm0.018,\quad g_T^d=-0.225\pm0.008,
\end{align}
with 90\% C.L. uncertainties. The values are evaluated at the scale of $4\,\rm GeV^2$.

To estimate the impact of these experiments, we still use the strange quark tensor charge from the lattice simulation~\cite{Bhattacharya:2015esa}. Following the same procedure, we obtain the constraint from proton and neutron EDM limits shown in Figure~\ref{nucleon_edm}, and the combination of proton and neutron results gives the upper limits on quark EDMs as
\begin{align}
|d_u|&\leq6.11\times10^{-25}\,e\cdot{\rm cm},\\
|d_d|&\leq9.70\times10^{-25}\,e\cdot{\rm cm},
\end{align}
at the scale of $4\,\rm GeV^2$. The final results that include the additional 10\% uncertainty of the isospin symmetry breaking effect are listed in Table~\ref{quark_edm}.

As observed from Figure~\ref{nucleon_edm}, although the precision of tensor charge extractions is improved by one order of magnitude, the impact on the constraint on quark EDMs is not significant. Therefore as expected, more sensitive proton and neutron EDM experiments are necessary.

The precision of the neutron EDM measurements has improved by six orders of magnitude since the first experiment by Purcell {\it et al.}~\cite{Purcell:1950zz,Smith:1957ht}. The goal of the next generation neutron EDM experiments~\cite{nEDMexps} is to further improve the sensitivity by two orders of magnitude. The statistical uncertainty of the measurement with ultra-cold neutrons depends on the electric field $E$, the number of neutrons $N$, and the storage time $\tau$ as~\cite{Golub:1994cg}
\begin{align}
\sigma\sim(E\sqrt{N\tau})^{-1}.
\end{align}
The approach~\cite{Golub:1994cg} that will be utilized in the next generation neutron EDM experiments~\cite{nEDMexps} will significantly increase $E$, $N$, and $\tau$, and will also have better control on systematic uncertainties. To estimate the impact of these experiments, we take the future neutron EDM limit as
\begin{align}
|d_n|\leq3.0\times10^{-28}\,e\cdot{\rm cm},
\end{align}
at 90\% confidence level. The result of its constraint on quark EDMs is shown in Figure~\ref{nucleon_edm}.

For the proton EDM measurement, a storage ring experiment~\cite{Anastassopoulos:2015ura} is proposed apart from the indirect measurements. The new method, which is based on the approach of minimizing the $g-2$ precession in the horizontal plane by using a radial electric field, can reach a sensitivity of $10^{-29}\,e\cdot{\rm cm}$ for the EDM measurement of the proton~\cite{Anastassopoulos:2015ura}. To estimate the impact of the experiment, we take the future proton EDM limit as
\begin{align}
|d_p|\leq2.6\times10^{-29}\,e\cdot{\rm cm},
\end{align}
at 90\% confidence level. The result of its constraint on quark EDMs is shown in Figure~\ref{nucleon_edm}.

Combining the results estimated with the precision of all these future experiments, we obtain the constraint on quark EDMs as
\begin{align}
|d_u|&\leq1.09\times10^{-27}\,e\cdot{\rm cm},\\
|d_d|&\leq6.53\times10^{-28}\,e\cdot{\rm cm},
\end{align}
at the scale of $4\,\rm GeV^2$. The limits that include the 10\% isospin symmetry breaking uncertainty are listed in Table~\ref{quark_edm}. The results of other combinations of current and future precisions of tensor charge and nucleon EDM measurements are also listed in Table~\ref{quark_edm}.

Compared with our current knowledge, the planned tensor charge and proton and neutron EDM measurements will improve the constraint on quark EDMs by three orders of magnitude.

\begin{table*}[htp]
\centering
\caption{Limits on quark EDMs. All EDM values are given as 90\% C.L. upper limits at the scale of $4\,\rm GeV^2$ in unit of $e\cdot{\rm cm}$. 10\% uncertainties are added to account for the isospin symmetry breaking. The current tensor charge precision refers to $g_T^{u}=0.405\pm0.130$ and $g_T^{d}=-0.225\pm0.092$, and the future tensor charge precision refers to $g_T^u=0.405\pm0.018$ and $g_T^d=-0.225\pm0.008$~\cite{Ye:2016prn}. The strange quark tensor charge $g_T^{s}=0.008\pm0.015$~\cite{Bhattacharya:2015esa} is used for both current and future tensor charge precisions. The current nucleon EDM limits refer to $|d_p|\leq2.6\times10^{-25}\,e\cdot{\rm cm}$ derived from the mercury atomic EDM measurement~\cite{Graner:2016ses} and $|d_n|\leq3.0\times10^{-26}\,e\cdot{\rm cm}$ from the neutron EDM measurement~\cite{Afach:2015sja}, and the future nucleon EDM limits refer to $|d_p|\leq2.6\times10^{-29}\,e\cdot{\rm cm}$ and $|d_n|\leq3.0\times10^{-28}\,e\cdot{\rm cm}$.}\label{quark_edm}
\begin{tabular}{ccccc}
\hline\hline
tensor charge precision & ~~~proton EDM limit~~~ & ~~~neutron EDM limit~~~ & ~~~~~~$d_u$ limit~~~~~~ & ~~~~~~$d_d$ limit~~~~~~\\
\hline
current & current & current & $1.27\times10^{-24}$ & $1.17\times10^{-24}$ \\
future & current & current & $6.72\times10^{-25}$ & $1.07\times10^{-24}$ \\
current & future & current & $1.27\times10^{-25}$ & $6.39\times10^{-26}$ \\
future & future & current & $1.14\times10^{-25}$ & $6.18\times10^{-26}$ \\
current & current & future & $1.16\times10^{-24}$ & $1.12\times10^{-24}$ \\
future & current & future & $5.60\times10^{-25}$ & $1.01\times10^{-24}$ \\
current & future & future & $1.36\times10^{-27}$ & $7.41\times10^{-28}$ \\
future & future & future & $1.20\times10^{-27}$ & $7.18\times10^{-28}$ \\
\hline\hline
\end{tabular}
\end{table*}

\section{The probe of new physics}

Since the SM CKM complex phase produces an extremely small quark EDM~\cite{Czarnecki:1997bu}, which can be viewed as a background within the experimental precisions at present and even in the next ten years, the quark EDM is one of the most sensitive probes of new physics models that provide additional $\cal CP$ violation sources. 

For new physics beyond the electroweak scale, the quark EDM is suppressed by the quark mass~\cite{Grzadkowski:2010es,DeRujula:1990db}. A simple dimensional analysis gives the quark EDM as $d_q\sim~e m_q/(4\pi\Lambda^2)$~\cite{Pospelov:2005pr}, where $\Lambda$ represents the scale of new physics. As presented in the previous section, the constraint on quark EDMs in the next ten years will be improved by three orders of magnitude. Therefore we expect it will be able to probe new physics at a scale $30\sim40$ times higher than that of its current reach. As a very rough estimation, we take light quark mass in~\cite{Olive:2016xmw} evaluated at $4\,\rm GeV^2$, and then  with the current quark EDM limit, which is about $10^{-24}\,e\cdot{\rm cm}$, the $\Lambda$ is about $1\,\rm TeV$. This energy scale is directly reached by the LHC. Future precise measurements of the tensor charge and the nucleon EDM will allow us to probe new physics up to $30\sim40\,\rm TeV$, which is above the LHC energy.

\begin{figure*}
\includegraphics[width=0.45\textwidth]{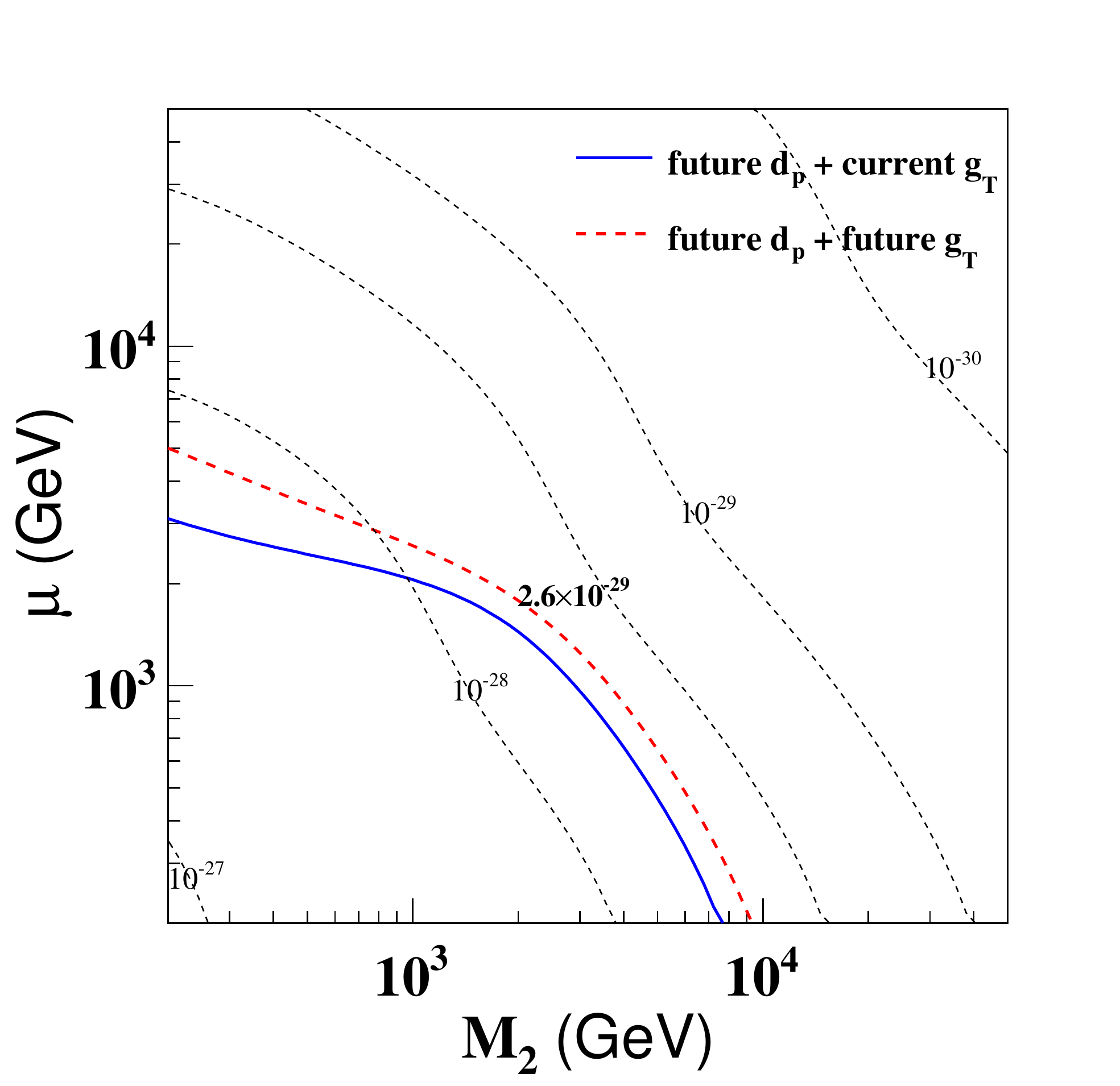}
d\includegraphics[width=0.45\textwidth]{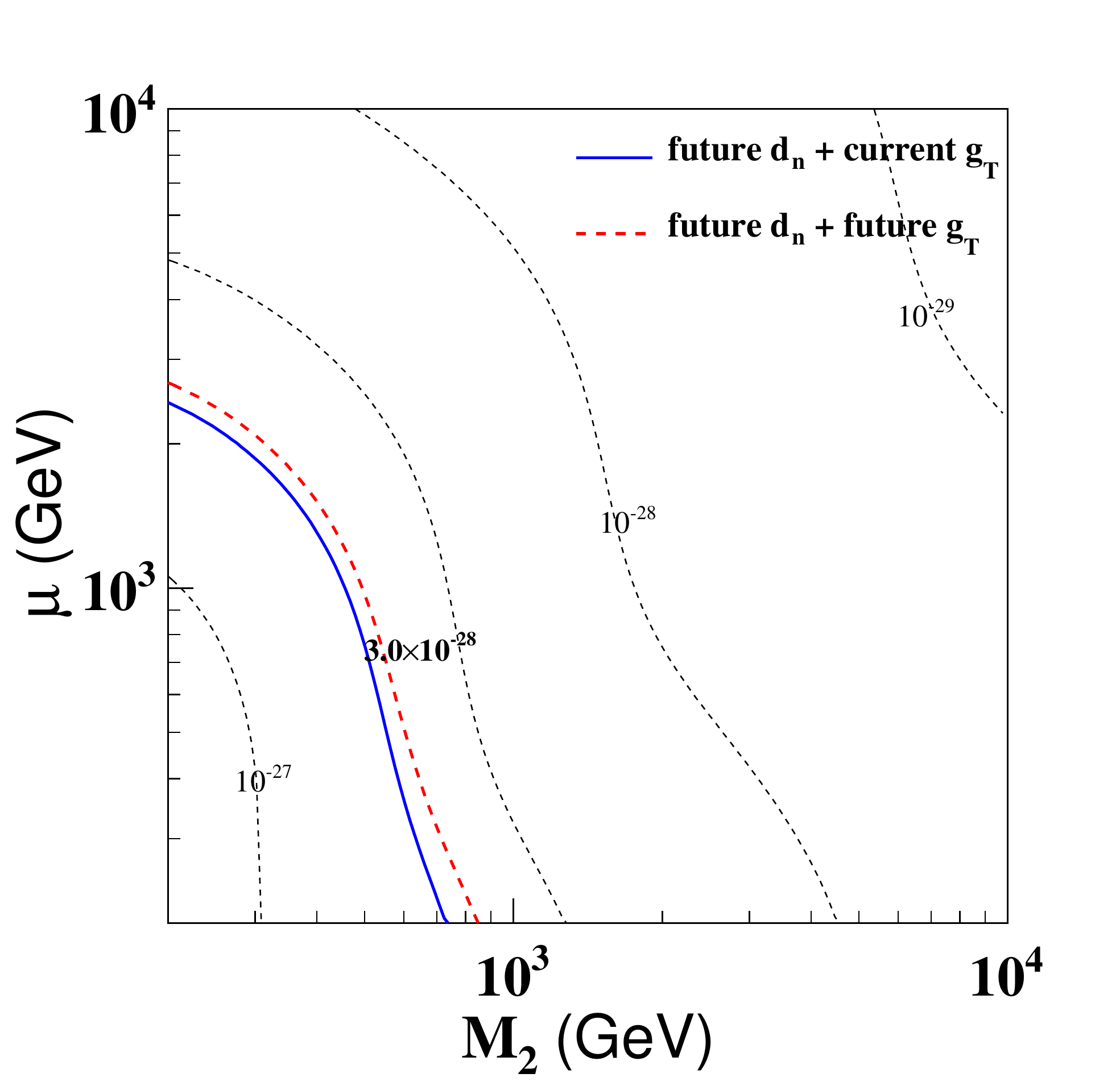}
\caption{The constraint on the split-supersymmetric model. The left panel is the constraint from the proton EDM, and the right panel is the constraint from the neutron EDM. The dotted (black) curves are constraints with precise tensor charge values, {\it i.e.} zero tensor charge uncertainties, the solid (blue) curves are those with current tensor charge uncertainties, and the dashed (red) curves are those with future tensor charge uncertainties. The nucleon EDM limit values marked on curves are in unit of $e\cdot{\rm cm}$. The area below the curves represents the excluded parameter space.}\label{split_susy}
\end{figure*}

As a specific example, we study the constraint on the parameter space of the split-supersymmetric extension of the SM, which is also investigated by the PNDME lattice QCD group~\cite{Bhattacharya:2015esa}. In the split-supersymmetric model~\cite{ArkaniHamed:2004fb}, one-loop contributions are highly suppressed by heavy sfermions~\cite{Giudice:2005rz,Li:2008kz} and thus it avoids the usual supersymmetric $\cal CP$ problem with current experimental limit on nucleon EDMs. In this model, the dominant contribution is from the quark EDM term arising from the two-loop level. With the theoretical calculation and the setup in Ref.~\cite{Giudice:2005rz}, namely the unified framework of gaugino masses at the grand unified theory scale and the sfermion mass of $10^9\,\rm GeV$, we estimate the constraints on $M_2$ and $\mu$, which are the mass parameters in the gaugino-higgsino sector. At the split limit with gaugino mass unification, quark EDMs are controlled by a single phase $\phi$ in an approximate linear way as $\sin\phi\sin2\beta$~\cite{Giudice:2005rz}, where $\tan\beta=v_u/v_d$ is the ratio between Higgs vacuum expectation values. In our estimation, we choose $\sin\phi=1$ and $\tan\beta=1$.  The results are shown in Figure~\ref{split_susy}. Our current limits have no sensitivity to this model. The comparison of the dotted (black) curves shows the impact of the improvement on nucleon EDM measurements, and the comparison between the solid (blue) and the dashed (red) curves shows the impact of the improvement on tensor charge extractions. So the combination of future nucleon EDM and tensor charge measurements will be a powerful tool for the search of beyond SM new physics and provide more stringent constraint on the parameter space of new physics models.

\section{Discussions and conclusions}

In this paper, we investigate the experimental constraint on quark EDMs by combining nucleon EDM measurements and tensor charge extractions. With the present sensitivity of the proton and neutron EDM experiments and the current precision of tensor charge extractions, we obtain the upper limit on quark EDMs as $1.27\times10^{-24}\,e\cdot{\rm cm}$ for the up quark and $1.17\times10^{-24}\,e\cdot{\rm cm}$ for the down quark at the scale of $4\,\rm GeV^2$. It corresponds to a probe of new physics roughly up to the energy scale of 1\,TeV, which is directly reached by the LHC. 

In the next ten years, both the sensitivity of nucleon EDM experiments and the precision of tensor charge extractions are expected to be dramatically improved. The planned SIDIS experiments at Jefferson Lab will improve the uncertainty of the determination of the tensor charge by one order of magnitude~\cite{Ye:2016prn}. The next generation neutron EDM experiments aim to improve the precision to $10^{-28}\,e\cdot{\rm cm}$~\cite{nEDMexps}. The proposed storage ring proton EDM experiment is capable of reaching a sensitivity of $10^{-29}\,e\cdot{\rm cm}$~\cite{Anastassopoulos:2015ura}. Our analysis shows that the combination of all these experiments is expected to improve the limit on quark EDMs by about three orders of magnitude. It means the energy scale of the quark EDM probe of new physics models will increase by $30\sim40$ times, and thus is above the LHC energy. Taking the split-supersymmetric model as an example, we show the impact of the improvements on nucleon EDM measurements and tensor extractions with the future experiments. 
Therefore it will become an important approach for us to explore the new source of $\cal CP$ violating effects and hence the baryogenesis mechanism of our universe.

Our analysis in this study is based on the sole contribution assumption. The strong $\cal CP$ violation $\theta$-term is set to zero. Other contributions such as the chromo-EDM and the Weinberg term are also neglected. In some model like the split-supersymmetric model we investigated, the quark EDM term dominates and one can neglect the chromo-EDM and the Weinberg term contributions at the leading order. In general, it is possible to have cancellations among different sources. We leave more complete investigations of all sources to future studies.

The isospin symmetry is another assumption we have used in our analysis. The 10\% uncertainty that we have added based on our empirical estimation is more or less arbitrary. With our current knowledge, we cannot quantify this uncertainty. Better understandings of the nucleon structure both theoretically and experimentally in the future may help us to have more accurate estimation of this uncertainty.

The validation of the neglect of the charm as well as heavier flavor contributions depends on the size of its tensor charge. Since the charm quark is about 600 times heavier than the up quark, if we require the charm quark contribution to be one order of magnitude smaller than the up quark contribution, the charm quark tensor charge is required to be smaller than $7\times10^{-5}$ at the scale of $4\,\rm GeV^2$. To our best knowledge, there are no theoretical calculations or experimental extractions of the charm quark tensor charge up to now. Future studies may tell us if the drop of heavy flavor terms is allowed.

In conclusion, the quark EDM is a sensitive probe to additional $\cal CP$ violation sources beyond the SM. The upcoming experiments will significantly improve the constraint on the quark EDM and hence make it a much more powerful tool to test the SM and to search for new physics models. 

\acknowledgments{
We are grateful to Kalyan Allada, Jian-Ping Chen, Bradley W. Filippone, Robert Golub, Zhong-Bo Kang, Alexei Prokudin, Nobuo Sato, Peng Sun, Zhihong Ye, and Feng Yuan for useful discussions.
This work is supported in part by U.S. Department of Energy under contract number DE-FG02-03ER41231, and by the Duke Kunshan University.
}

\end{document}